# Crossover from lamellar to spongy ice morphologies within a single ice crystal during unidirectional freezing of an aqueous solution

*Tongxin Zhang, Zhijun Wang\*, Lilin Wang\*, Junjie Li, Jincheng Wang*

*State Key Laboratory of Solidification Processing, Northwestern Polytechnical University, Xi'an 710072, China*

**Abstract:** Ice growth from liquid phase has been extensively investigated in various conditions, especially for ice freely grown in undercooled water and aqueous solutions. Although unidirectional ice growth plays a significant role in sea ice and freeze casting, the detailed pattern formation of unidirectionally grown ice in an aqueous solution remains elusive. For the first time, we in situ proved a crossover from lamellar to spongy ice morphologies of a single ice crystal via unidirectional freezing of an aqueous solution. The spongy ice morphology originates from the intersect of tilted lamellar ice and is observed in a single ice crystal, which is intrinsically different from the competitive growth of bi-crystal composed of two differently orientated grains in directional solidification. These results provide a complete physical picture of unidirectionally grown ice from aqueous solution and are believed to promote our understanding of various pattern of ice in many relevant domains where pattern formation of ice crystal is vital.

Motivated by the vast engineering applications in many communities such as pharmaceutics [1], cryobiology [2-7], material sciences [8-12] and cold region investigations concerning ice accretion [13-17], as well as by fundamental scientific curiosity of ice as an excellent testing ground for investigations of faceted crystal, the microstructure features of ice growth from liquid phase have been intensively investigated both experimentally [18-22] and computationally [23-25]. Numerous previous reports on ice growth in both free growth and unidirectional growth provided useful clues to imagine the physical picture of ice morphology in liquid phase, which consists of lamellar ice [26-30] or "spongy ice" of crossed ice lamellae with a portion

\* Corresponding author. zhjwang@nwpu.edu.cn
\* Corresponding author. wlilin@nwpu.edu.cn

of homogeneously distributed brine pockets [31-33].

Ice is well-known as a highly anisotropic inorganic crystal both kinetically and thermodynamically, and its growth is intriguing but complex. Freely grown ice from liquid phase is a typical example of such complexity, which has been widely reported to experience a morphological transition in its melt at a specific undercooling, during which the coplanar growth along basal plane of ice will become non-coplanar and form hexagonal pyramid pairs [34, 35]. The growth habit of freely grown ice in aqueous solutions were similar [36, 37], and side-branches were much easier to occur due to diffusion-controlled instability. In contrast to the numerous reports on freely grown ice, our understanding of morphogenesis of unidirectionally grown ice crystal is quite limited. In a simple analogy, arrays of ice lamellae occurred in unidirectional growth may also become non-coplanar at some growth conditions.

Nevertheless, many unidirectional freezing investigations only provided the formation of lamellar ice. These previous reports are limited in addressing the complex freezing pattern of ice from liquid phase. Besides, tilting of lamellar ice from the direction of thermal gradient was usually claimed to be a competition between unparalleled thermal gradient and preferred crystal orientation (the direction parallel to basal plane of ice) [38, 39], which is widely applied in solidification of metals. However, this type of mechanism holds valid for ice only when thermal gradient is not parallel to the preferred orientation of ice. And there is an obvious incompletion in this explanation, given the previously introduced crossover from coplanar to non-coplanar growth of ice from liquid phase. Therefore, it is important to obtain a complete physical picture of morphological evolution of ice around this crossover in a unidirectional freezing manner within a single ice crystal of appropriate orientation.

The morphological evolution of a unidirectionally grown single ice crystal can play a unique role as an exploratory tool for elucidating the various pattern of ice growth in many relevant domains. Therefore, it is the principal aim of this Letter to answer the question of whether a transition from coplanar to non-coplanar growth exists in directional solidification via well-designed experiments within a single ice crystal of controlled crystal orientation.

Based on previous reports, ice growth habit in its pure melt or aqueous solution refers to its C axis and basal plane {0001}. **Figure 1** shows a series of morphological transition of freely grown ice with increasing bath undercooling and relevant explanations, where the C axis (red solid rod) and basal plane {0001} (black dotted square) are marked. When the bath undercooling $\Delta T_{bath}$ is lower than around 2.7 K [35, 40], the coplanar growth along basal plane of ice will grow like a circular disk (see **Fig. 1 (a)**) or "plane stellar dendrites" [36] (see **Fig. 1 (b)**) at higher $\Delta T_{bath}$. When $\Delta T_{bath}$ exceeds a critical value of around 2.7 K, the coplanar growth along basal plane of ice will become non-coplanar and form hexagonal pyramid pairs with an open angle (see **Fig. 1 (c)**) which depends on the magnitude of $\Delta T_{bath}$ [41] and added solutes [18, 42]. And further increment of $\Delta T_{bath}$ will produce more complex side-branches along the primary arms (see **Fig. 1 (d)**) [43].

The transition from coplanar to non-coplanar growth is previously addressed by a physical picture which claimed that the actual ice tip velocity was determined by the normal velocities in both basal and edge planes [34]. In the coplanar growth regime, freely grown ice usually appears like a flat disk lying in the basal plane of ice. The normal growth velocity of basal plane is limited by two-dimensional nucleation, while the planes perpendicular to the basal plane grows in a continuous manner [44] without the limitation of nucleation process [45] as shown in **Fig. 1 (e)**. In the non-coplanar growth regime [35, 40], the ice disk no longer grows within the basal plane but forms three-dimensional microstructure of a twelve-sided double pyramid as shown in **Fig. 1 (f)**. Levi et al. [46, 47] attributed this transition to the altered interface roughness of basal plane of ice. The possible change of the smoothness of basal plane of ice may be further related to the concept of "roughening transition" [48], which claims that facets on the S/L interface can be destroyed and become rough by larger growth velocity. Braslavsky et al. [35] discussed the continuous tilting of tip of freely grown ice against various bath undercooling in the light of microscopic solvability combined with the behavior of the surface tension and the kinetic effect as a function of crystallographic orientation, yet this phenomenon remains elusive. Further investigations are needed based on the anisotropy of ice in its growth kinetics and/or interfacial energy and possible roughening transition of basal plane of ice.

Inspired by the reported growth morphology of freely grown ice in **Fig. 1**, there is a possible picture of morphological transition for unidirectionally grown ice in thin rectangular capillary as shown in **Fig. 2**. The crystal orientation of ice in capillary for unidirectional freezing of an aqueous solution is supposed to satisfy the relation of $\vec{V}_p \parallel \vec{G} \parallel \{0001\} \parallel \vec{L}$ as indicated in **Fig. 2**, where $\vec{V}_p$, $\vec{G}$, $\{0001\}$ and $\vec{L}$ represent pulling velocity, thermal gradient, basal plane of ice and incident light (indicated by a solid circle with a tilted cross inside in red), respectively. It is expected in **Fig. 2 (a)** that when pulling velocity of unidirectional freezing is low enough, the tip undercooling of lamellar ice is also relatively low, and tilting of cells due to the transition from coplanar to non-coplanar will not occur. And lamellar ice with cellular tip morphology is expected to occur after planar instability. As later shown in **Fig. 2 (b)**, when the pulling velocity becomes larger, tilting of lamellar ice due to the transition from coplanar to non-coplanar growth is expected to occur, and two families of lamellar ice with different tilting directions should appear in different positions perpendicular to the pulling velocity. Each family of ice lamellae occupies the entire thickness of the capillary, resulting the expected "tilted lamellar ice" with a certain tilting angle. In **Fig. 2 (c)**, when the pulling velocity further increases, tilted lamellar ice with finer ice branches are expected to occupy a portion of thickness of the capillary and form "spongy ice" consisting of overlapped ice branches with trapped brine pockets of rhombus shape.

There should be an emphasis on **Fig. 2 (b)** that, one is very likely to mistakenly interpret the ice growth morphology in **Fig. 2 (b)** as a case of bi-crystal competition of two ice grains with different orientations [49, 50], which, however, is exactly the ice morphology within a single ice crystal with an orientation of $\vec{V}_p \parallel \vec{G} \parallel \{0001\} \parallel \vec{L}$ in **Fig. 2**. The physical pictures of bi-crystal competition and the experimental observations in this Letter are intrinsically different.

In order to confirm the speculation of morphological transition in **Fig. 2**, dozens of in situ experiments were conducted with the aid of a customized horizontal Bridgeman solidification platform. The detailed information of our experimental apparatus in this Letter has been described elsewhere [51-53]. Ultrapure water (18.25 MΩ) was chosen as the solvent for the preparations of aqueous solution. KCl (AR, 99.5 %) was applied as the solute. KCl solution with a concentration of 0.1 M was prepared at room temperature by dissolving KCl powder in ultrapure water. Prior to freezing experiments, KCl solution was degassed under vacuum condition for half an hour to prevent possible occurrence of air bubbles near the S/L interface during freezing. Rectangular glass capillary with an inner space dimension of 0.10 x 2.00 mm$^2$ (VitroCom brand) was adopted as sample cell for unidirectional freezing. The freezing experiments were conducted under varying pulling velocities approximately ranging from 3.57 um/s to 58.82 um/s. It should be noted that each freezing sample was kept static for a time interval of more than one hour to make the S/L interface planar before next used pulling velocity so as to eliminate the effect of ice morphology formed in every previous freezing process. Since this Letter deals with the crossover from coplanar to non-coplanar growth, the direction of C axis is specially manipulated in our experiments in case non-coplanar growth can be observed in the field of our microscope upon its occurrence. The single ice crystal in KCl solution was then manipulated to have an orientation relation of $\vec{V}_P \parallel \vec{G} \parallel \{0001\} \parallel \vec{L}$ on our freezing stage by means of a step-by-step methodology reported elsewhere [54]. The ice crystal orientation is simultaneously detected through a pair of polarizers to guarantee that the crystal orientation remained unchanged during each run of unidirectional freezing experiments.

**Figure 3** shows the most important morphological information from the experiments. The morphology of single ice crystal varies significantly with increased pulling velocity, showing vividly the occurrence of three typical morphologies: **(a)** "lamellar ice", **(b)** "tilted lamellar ice" and **(c)** "spongy ice". The results in **Fig. 3 (a-c)** are in qualitative agreement with the speculation in **Fig. 2 (a-c)**. And these results validate a crossover from "lamellar ice" to "spongy ice" morphologies under the precondition of paralleled thermal gradient with respect to the preferred orientation of a single ice crystal. Interestingly, ice tip morphology also varies with increased pulling velocity, from rounded cellular ice tip (**Fig. 3 (a)**) to knife-edged ice tip (**Fig. 3 (b-c)**) with a smooth faceted edge on one side of the lamellar ice stacked side by side. The interesting morphology of "tilted lamellar ice" in **Fig. 3 (b)** is attributed to the non-planar growth of hexagonal pyramid pairs in ice which is previously mentioned.

During many runs of unidirectional freezing experiments, there is also an interesting phenomenon of overgrown between two families of "tilted lamellar ice" that are "convergent" (indicated by the red solid arrows tilting up and down) as shown in **Fig. 4**. The solid circles plotted in **Fig. 4** clearly reveal how two ice lamellae that are tilting-down (indicated by the red solid arrow tilting down) are gradually overgrown by their neighboring ice lamellae that are tilting-up (indicated by the red solid arrow tilting up) in a time interval of about three minutes (from **Fig. 4 (a)** 00 h:47 min: 20 s to **Fig. 4 (c)** 00 h: 50 min: 55 s). Such overgrown or competition growth phenomenon can only be observed at moderately high pulling velocity when each family of tilted lamellar ice occupies the whole thickness of the capillary. As time goes by, the overgrown process will gradually eliminate convergent lamellar ice microstructure. And the tilted lamellar ice morphology that remain stable usually consists of either only one family of tilted lamellar ice. For example, in different runs of the experiments, **Fig. 5 (a)** and **Fig. 5 (b)** show stable tilting-down and tilting-up lamellar microstructure (indicated by the red solid arrows tilting down and/or up) after their overgrown processes. In **Fig. 5 (c)**, two families of tilted lamellar ice that are "divergent" (indicated by two red solid arrows tilting down and up) can remain stable after an overgrown process. When the pulling velocity further increases to some extent, two families of tilted lamellar ice can coexist with each family of thinner ice lamellae occupying half of thickness of the capillary and form "spongy ice" as speculated in **Fig. 2 (c)** and observed experimentally in **Fig. 3 (c)**.

In conclusion, this Letter provides an investigation on crossover from "lamellar ice" to "spongy ice" morphologies within a single ice crystal via unidirectional freezing of an aqueous solution with manipulated ice orientation. The experimental results are in qualitative agreement with the speculation based on previous reports concerning morphogenesis of freely grown ice. Similar to the morphology for freely grown ice, the unidirectionally grown ice in aqueous KCl solution produces more complex patterns with a similar crossover from coplanar to non-coplanar growth even when the preferred crystal orientation is set parallel to thermal gradient prior to unidirectional freezing. The overgrown phenomenon among families of tilted lamellar ice is also observed within a single ice crystal for the first time. These results are suggested to deepen our understanding of various pattern of ice during unidirectional freezing of more sophisticated systems such as polymer solutions and colloidal suspensions. And the complete physical picture of unidirectionally grown ice proposed in this Letter is believed to be of great significance in many relevant domains where pattern formation of ice plays a significant role.

**Acknowledgements**

This work was supported by the National Key R&D Program of China (Grant No.2018YFB1106003), National Natural Science Foundation of China (Grant No. 51701155), the Research Fund of the State Key Laboratory of Solidification

Processing (NPU), China (Grant No. 2020-TS-06), and the Fundamental Research Funds for the Central Universities (3102019ZD0402).

[52] T. Zhang, L. Wang, Z. Wang, J. Li, J. Wang, Single ice crystal growth with controlled orientation during directional freezing, The Journal of Physical Chemistry B, 2020 (accepted).

[53] T. Zhang, Z. Wang, L. Wang, J. Li, J. Wang, The planar interface instability during freezing of a polymer solution: Diffusion-controlled or not?, on arXiv:2012.04274v2, 2020.

[54] T.X. Zhang, Z.J. Wang, L.L. Wang, J.J. Li, X. Lin, J.C. Wang, Orientation determination and manipulation of single ice crystal via unidirectional solidification, Acta Physica Sinica, 67 (2018).
10

**Figure and table captions:**

FIG. 1 The schematic diagram of growth morphology of freely grown single ice crystal in its pure melt as a function of bath undercooling of up to 10 K and relevant explanations. The crystal orientation of ice is indicated by C axis and/or basal plane {0001} of ice. With the increase of bath undercooling, the ice morphology is reported to experience a series of transition of **(a)** "circular ice disk" (0 K< $\Delta T_{bath}$ < 0.9 K) — **(b)** "plane stellar dendrites" (0.9 K< $\Delta T_{bath}$ < 2.7 K) — **(c)** "simple double pyramids" with an open angle (2.7 K< $\Delta T_{bath}$ < 5.5 K) — **(d)** "complex double pyramids with side-branches" (5.5 K< $\Delta T_{bath}$ < 10 K). The critical bath undercooling of crossover from coplanar to non-coplanar growth is reported to be roughly 2.7 K as indicated in the figure. The morphology in **(a)** and **(b)** is addressed by anisotropy in growth kinetics and/or interfacial energy of basal and edge planes of ice as shown in **(e)**. The morphology in **(c)** and **(d)** is addressed by a combined effect of comparable growth kinetics of basal and edge planes at the lamellar ice tip as shown in **(f)**.

FIG. 2 Proposed morphological transition of unidirectional grown ice in 0.1M KCl solution against increased pulling velocity. The crystal orientation of ice in capillary for unidirectional freezing is supposed to be the same, satisfying the relation of $\vec{V}_p \parallel \vec{G} \parallel \{0001\} \parallel \vec{L}$ indicated by the solid arrow at the bottom of the figure ($\vec{L}$ is the direction of incident light). With the increase of pulling velocity, the ice morphology is suggested to experience a transition of **(a)** "lamellar ice"—**(b)** "tilted lamellar ice" (open angle is indicated by two red solid arrows and a segmental arc between them)— **(c)** "spongy ice" (open angle is indicated by two red solid arrows and a segmental arc between them).

**FIG. 3** In situ observed morphological transition of unidirectional grown ice in 0.1M KCl solution in this paper. The crystal orientation of ice in capillary for unidirectional freezing is manipulated to be the same, satisfying the relation of $\vec{V}_p \parallel \vec{G} \parallel \{0001\} \parallel \vec{L}$ indicated by the solid arrow at the bottom of the figure ($\vec{L}$ is the direction of incident light). With the increase of pulling velocity, the ice morphology is confirmed to experience a transition of **(a)** "horizontal cellular ice" ($V_p$ = 3.57 um/s and $G$ = 2.72 K/mm) — **(b)** "tilted dendritic ice" ($V_p$ = 13.17 um/s and $G$ = 2.72 K/mm, open angle is indicated by two red solid arrows and a segmental arc between them) — **(c)** "brine spongy ice" ($V_p$ = 58.82 um/s and $G$ = 1.34 K/mm, open angle is indicated by two red solid arrows and a segmental arc between them). The scale bar in each figure is 250 um.

**FIG. 4** In situ observed overgrown phenomenon of one family of tilted cells/dendrites that are convergent with respect to another family with a different tilting direction. As time goes by from time position of **(a)** 00 hour: 47 min: 20 s, **(b)** 00 hour: 50 min: 05 s to **(c)** 00 hour: 50 min: 55 s, overgrown between convergent families of tilted cells/dendrites occurs in the areas within the red solid circles in **(a-c)** ($V_p$ = 11.8 um/s and $G$ = 2.71 K/mm). Growth directions of ice lamellar tips are indicated by the red solid arrows. The crystal orientation of ice in capillary for unidirectional freezing is manipulated to be the same, satisfying the relation of $\vec{V}_p \parallel \vec{G} \parallel \{0001\} \parallel \vec{L}$ ($\vec{L}$ is the direction of incident light) indicated by the solid arrow at the bottom of the figure. The scale bar in each figure is 250 um.

**FIG. 5** Three types of in situ observed morphology of tilted cells/dendrites under steady state unidirectional freezing. **(a)** Tilting-down lamellar microstructure ($V_p$ = 11.80 um/s and $G$ = 2.42 K/mm). **(b)** Tilting-up lamellar microstructure ($V_p$ = 11.70 um/s and $G$ = 2.42 K/mm). **(c)** Divergent lamellar microstructure ($V_p$ = 13.17 um/s and $G$ = 2.42 K/mm). Growth directions of ice lamellar tips are indicated by the red

solid arrows. The crystal orientation of ice in capillary for unidirectional freezing is manipulated to be the same, satisfying the relation of $\vec{V}_P \parallel \vec{G} \parallel \{0001\} \parallel \vec{L}$ indicated by the solid arrow at the bottom of the figure ($\vec{L}$ is the direction of incident light). The scale bar in each figure is 250 um.

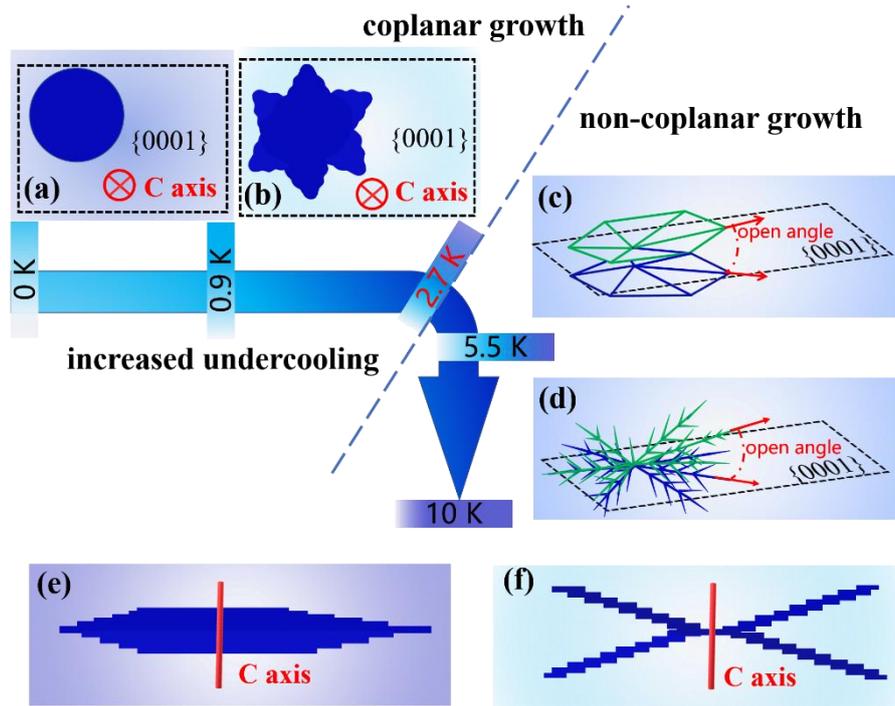

**FIG. 1** The schematic diagram of growth morphology of freely grown single ice crystal in its pure melt as a function of bath undercooling of up to 10 K and relevant explanations. The crystal orientation of ice is indicated by C axis and/or basal plane {0001} of ice. With the increase of bath undercooling, the ice morphology is reported to experience a series of transition of **(a)** "circular ice disk" (0 K< $\Delta T_{bath}$ < 0.9 K) — **(b)** "plane stellar dendrites" (0.9 K< $\Delta T_{bath}$ < 2.7 K) — **(c)** "simple double pyramids" with an open angle (2.7 K< $\Delta T_{bath}$ < 5.5 K) — **(d)** "complex double pyramids with side-branches" (5.5 K< $\Delta T_{bath}$ < 10 K). The critical bath undercooling of crossover from coplanar to non-coplanar growth is reported to be roughly 2.7 K as indicated in the figure. The morphology in **(a)** and **(b)** is addressed by anisotropy in growth kinetics and/or interfacial energy of basal and edge planes of ice as shown in **(e)**. The morphology in **(c)** and **(d)** is addressed by a combined effect of comparable growth kinetics of basal and edge planes at the lamellar ice tip as shown in **(f)**.

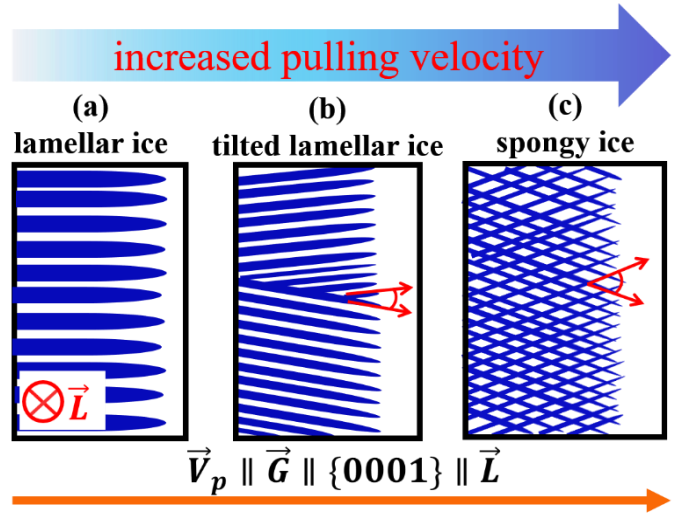

**FIG. 2** Proposed morphological transition of unidirectional grown ice in 0.1M KCl solution against increased pulling velocity. The crystal orientation of ice in capillary for unidirectional freezing is supposed to be the same, satisfying the relation of $\vec{V}_p \parallel \vec{G} \parallel \{0001\} \parallel \vec{L}$ indicated by the solid arrow at the bottom of the figure ($\vec{L}$ is the direction of incident light). With the increase of pulling velocity, the ice morphology is suggested to experience a transition of **(a)** "lamellar ice"—**(b)** "tilted lamellar ice" (open angle is indicated by two red solid arrows and a segmental arc between them)—**(c)** "spongy ice" (open angle is indicated by two red solid arrows and a segmental arc between them).

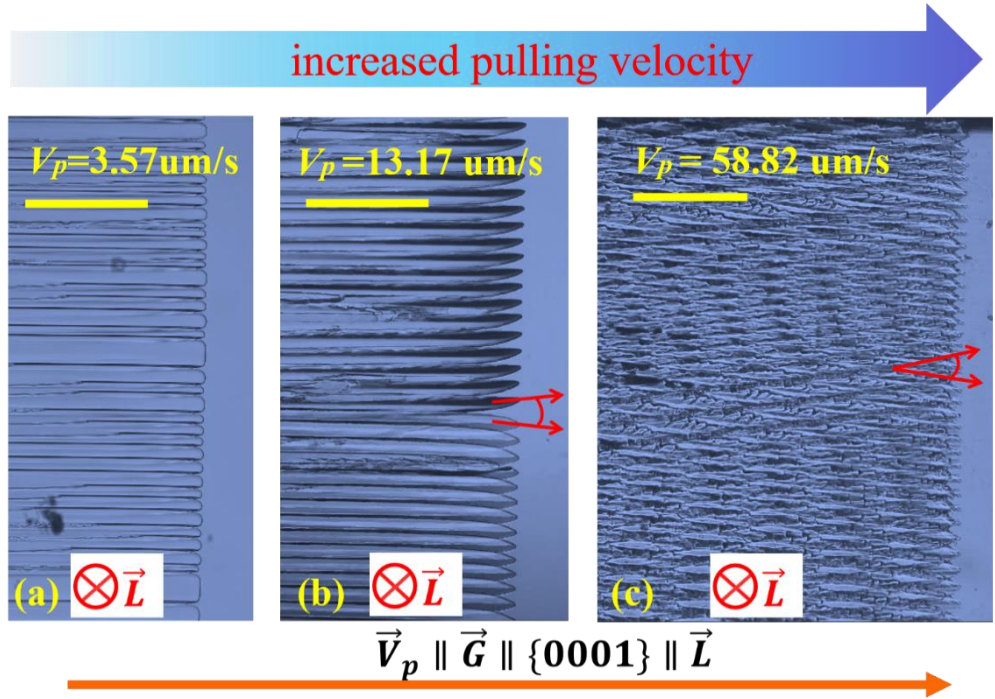

**FIG. 3** In situ observed morphological transition of unidirectional grown ice in 0.1M KCl solution in this paper. The crystal orientation of ice in capillary for unidirectional freezing is manipulated to be the same, satisfying the relation of $\vec{V}_p \parallel \vec{G} \parallel \{0001\} \parallel \vec{L}$ indicated by the solid arrow at the bottom of the figure ($\vec{L}$ is the direction of incident light). With the increase of pulling velocity, the ice morphology is confirmed to experience a transition of **(a)** "horizontal cellular ice" ($V_p$ = 3.57 um/s and $G$ = 2.72 K/mm) — **(b)** "tilted dendritic ice" ($V_p$ = 13.17 um/s and $G$ = 2.72 K/mm, open angle is indicated by two red solid arrows and a segmental arc between them) — **(c)** "brine spongy ice" ($V_p$ = 58.82 um/s and $G$ = 1.34 K/mm, open angle is indicated by two red solid arrows and a segmental arc between them). The scale bar in each figure is 250 um.

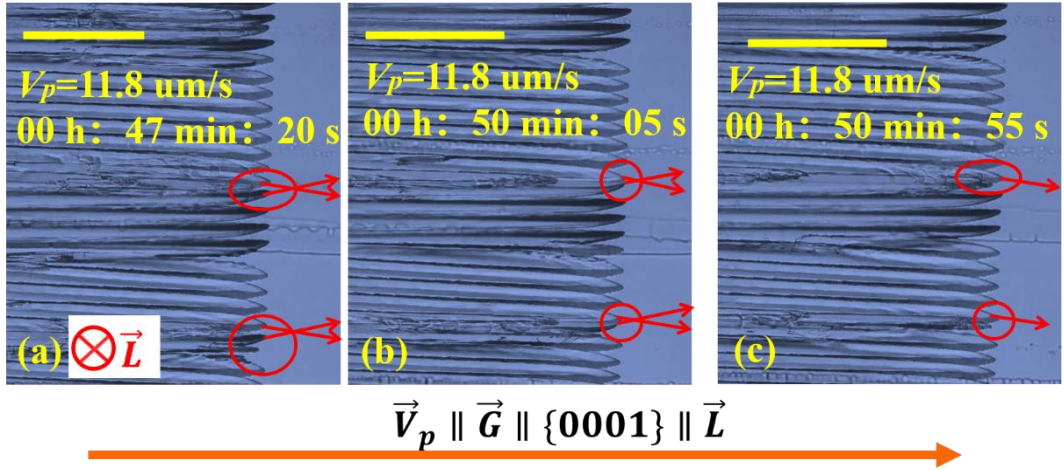

**FIG. 4** In situ observed overgrown phenomenon of one family of tilted cells/dendrites that are convergent with respect to another family with a different tilting direction. As time goes by from time position of **(a)** 00 hour: 47 min: 20 s, **(b)** 00 hour: 50 min: 05 s to **(c)** 00 hour: 50 min: 55 s, overgrown between convergent families of tilted cells/dendrites occurs in the areas within the red solid circles in **(a-c)** ($V_p$ = 11.8 um/s and $G$ = 2.71 K/mm). Growth directions of ice lamellar tips are indicated by the red solid arrows. The crystal orientation of ice in capillary for unidirectional freezing is manipulated to be the same, satisfying the relation of $\vec{V}_p \parallel \vec{G} \parallel \{0001\} \parallel \vec{L}$ ($\vec{L}$ is the direction of incident light) indicated by the solid arrow at the bottom of the figure. The scale bar in each figure is 250 um.

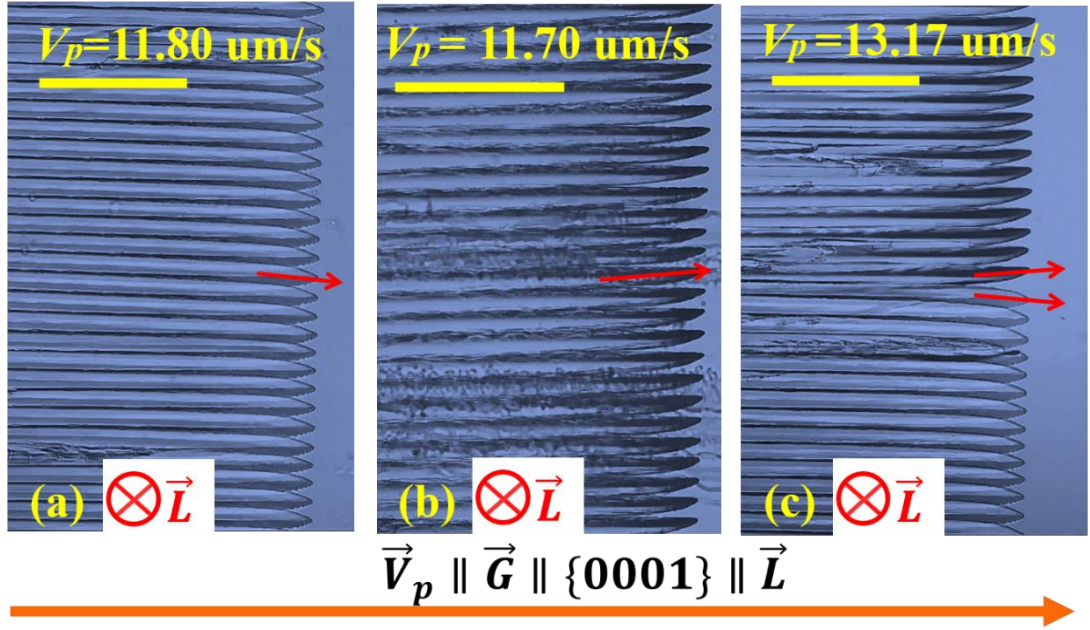

**FIG. 5** Three types of in situ observed morphology of tilted cells/dendrites under steady state unidirectional freezing. **(a)** Tilting-down lamellar microstructure ($V_p$ = 11.80 um/s and $G$ = 2.42 K/mm). **(b)** Tilting-up lamellar microstructure ($V_p$ = 11.70 um/s and $G$ = 2.42 K/mm). **(c)** Divergent lamellar microstructure ($V_p$ = 13.17 um/s and $G$ = 2.42 K/mm). Growth directions of ice lamellar tips are indicated by the red solid arrows. The crystal orientation of ice in capillary for unidirectional freezing is manipulated to be the same, satisfying the relation of $\vec{V}_p \parallel \vec{G} \parallel \{0001\} \parallel \vec{L}$ indicated by the solid arrow at the bottom of the figure ($\vec{L}$ is the direction of incident light). The scale bar in each figure is 250 um.